\documentclass{article}
\usepackage{spconf,amsmath,graphicx}
\usepackage{multirow}
\usepackage{bm}
\usepackage{cite}
\usepackage{xcolor}

\newcommand{\bhline}[1]{\noalign{\hrule height #1}} 

\title{Selecting N-lowest scores for training MOS prediction models}
\name{Yuto Kondo, Hirokazu Kameoka, Kou Tanaka, Takuhiro Kaneko\thanks{This
work was supported by JST CREST Grant Number JP-MJCR19A3, Japan.}}
\address{NTT Corporation, Japan}
\begin{document}
\ninept

\maketitle
\begin{abstract}
The automatic speech quality assessment (SQA) has been extensively studied to predict the speech quality without time-consuming questionnaires. Recently, neural-based SQA models have been actively developed for speech samples produced by text-to-speech or voice conversion, with a primary focus on training mean opinion score (MOS) prediction models. The quality of each speech sample may not be consistent across the entire duration, and it remains unclear which segments of the speech receive the primary focus from humans when assigning subjective evaluation for MOS calculation.
We hypothesize that when humans rate speech, they tend to assign more weight to low-quality speech segments, and the variance in ratings for each sample is mainly due to accidental assignment of higher scores when overlooking the poor quality speech segments. Motivated by the hypothesis, we analyze the VCC2018 and BVCC datasets.
Based on the hypothesis, we propose the more reliable representative value $N_{\rm low}$-MOS, the mean of the $N$-lowest opinion scores. Our experiments show that LCC and SRCC improve compared to regular MOS when employing $N_{\rm low}$-MOS to MOSNet training. This result suggests that $N_{\rm low}$-MOS is a more intrinsic representative value of subjective speech quality and makes MOSNet a better comparator of VC models.
\end{abstract}
\section{Introduction}
\label{sec:intro}
\vspace*{-0.2cm}
The automatic speech quality assessment (SQA) has been extensively studied with the aim of predicting the speech quality of acoustic equipment or speech processing systems without time-consuming questionnaires. 
For a long time, many objective metrics for the speech quality have been developed, limited to specific speech applications~\cite{ITU-R1387,rix2001perceptual,kim2005anique,falk2010non,beerends2013perceptual,chinen2020visqol}. For instance, PESQ~\cite{rix2001perceptual} is designed to evaluate telephone speech using not only the evaluated speech but a corresponding clean reference speech.
In recent years, for assessing speech signals generated by text-to-speech (TTS)~\cite{ren2020fastspeech,popov2021grad} or voice conversion (VC)~\cite{qian2019autovc,kaneko2019stargan,popov2021diffusion}, for which such clean reference speech may not be prepared, many neural-based automatic SQA models without the use of such a reference are proposed~\cite{patton2016automos,fu2018quality,lo2019mosnet,manocha2022speech, leng2021mbnet,huang2022ldnet}. Most of the models, e.g., MOSNet~\cite{lo2019mosnet}, are trained to predict the mean opinion score (MOS)~\cite{ITU-TP800} at the utterance-level, where MOS is a long-standing representative value of subjective speech quality. MOS is calculated by averaging all ratings labeled to a speech sample with absolute category rating (ACR) format~\cite{ITU-TP800} (e.g., five point opinion scale which consists of ``1: bad," ``2: poor," ``3: fair," ``4: good," and ``5: excellent.").
\begin{figure}[t]
 \vspace*{0cm}
 \centering
 \hspace*{-0cm}
\includegraphics[width=1.00\columnwidth]{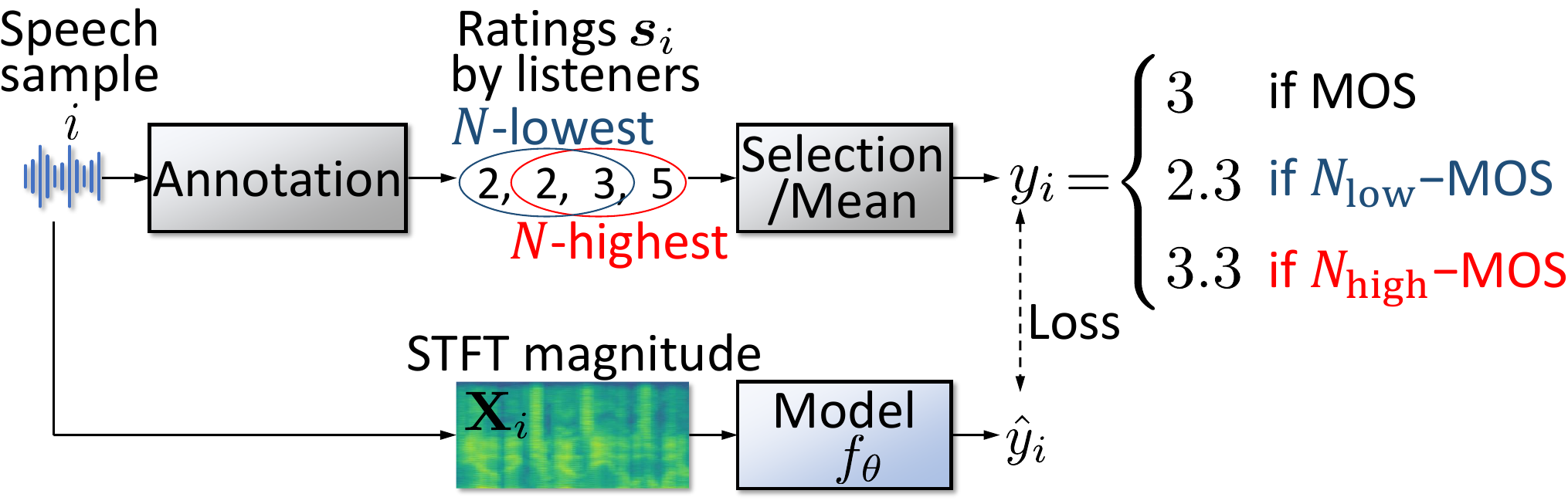}
\vspace*{-0.8cm}
\caption{Model training using regular MOS and proposed $N_{\rm low}$-MOS. We also display $N_{\rm high}$-MOS, which is compared with $N_{\rm low}$-MOS in the experiment section. This figure corresponds to the case of $N=3$.}
 \label{fig:LMOS}
 \vspace*{-0.5cm}
\end{figure}

The subjective quality of a TTS speech sample may not be consistent across the entire duration, and it is not clear which speech segments are the main focus when assigning each subjective evaluation score for MOS calculation. We hypothesize that when humans subjectively rate speech, they tend to assign more weight to low-quality speech segments, and the variance in ratings for each speech sample is primarily attributed to accidental inclusion of higher scores due to overlooking or neglecting the poor quality speech segments. Though it is not easy to give a proof to the hypothesis perfectly, motivated by the hypothesis, we analyze the VCC2018~\cite{lorenzo_vcc2018} and BVCC~\cite{cooper2021voices} datasets.

From the above hypothesis, the application of MOS to subjective speech quality predictor models is consider to set the upper limit on the prediction performance, where the performance is measured by some metrics such as mean square error (MSE), linear correlation coefficient (LCC)~\cite{pearson1920notes_MOSNet}, and Spearman’s rank correlation coefficient (SRCC)~\cite{spearman1961proof_MOSNet}. We propose the more intrinsic representative value of subjective speech quality $N$-lowest MOS (or $N_{\rm low}$-MOS for short), the mean of the $N$-lowest opinion scores, and applying $N_{\rm low}$-MOS to training of subjective speech quality prediction models. Figure~\ref{fig:LMOS} shows the overview of the model training. 
We experimentally show that when employing $N_{\rm low}$-MOS to the MOSNet architecture, LCC and SRCC improve compared to regular MOS. This result suggests that $N_{\rm low}$-MOS is a more intrinsic representative value of subjective speech quality and more useful when comparing the quality of TTS or VC speech samples.

The main contributions of this paper are as follows:
\vspace*{-0.00cm}
\begin{itemize}
\setlength{\itemsep}{-0.0cm}
  \item We hypothesize that when humans subjectively rate speech, they tend to assign more weight to low-quality speech segments, and the variance in ratings for each speech sample is mainly due to accidental assignment of higher scores when overlooking the poor quality segments.
  \item  We analyze the VCC2018 and BVCC datasets, and for the first time discover that the distribution of all ratings for each speech sample tends to skew to the right.
  \item We propose the more intrinsic representative value of subjective speech quality $N_{\rm low}$-MOS, and applying $N_{\rm low}$-MOS to training of subjective speech quality prediction models in order to relax the limit of the prediction performance due to regular MOS.
  \item We experimentally confirm that LCC and SRCC improve compared to regular MOS when employing $N_{\rm low}$-MOS. This result suggests that $N_{\rm low}$-MOS is a more intrinsic representative value of subjective speech quality and more useful when comparing the quality of TTS or VC speech samples.
\end{itemize}
\vspace*{-0.0cm}

The rest of this paper is as follows:
In Section 2, we discuss related work.
Section 3 describes the VCC2018 and BVCC datasets, and analyze these datasets.
In Section 4, we propose $N_{\rm low}$-MOS and applying $N_{\rm low}$-MOS to prediction models. 
Section 5 presents the conditions and results of experiments.
Finally, we conclude this paper in Section 6.
\vspace*{-0.2cm}
\section{Related work}
\label{sec:intro}
\vspace*{-0.2cm}
\subsection{MOSNet~\cite{lo2019mosnet}}
\begin{figure}[t]
 \vspace*{0cm}
 \centering
 \hspace*{-0cm}
\includegraphics[width=0.98\columnwidth]{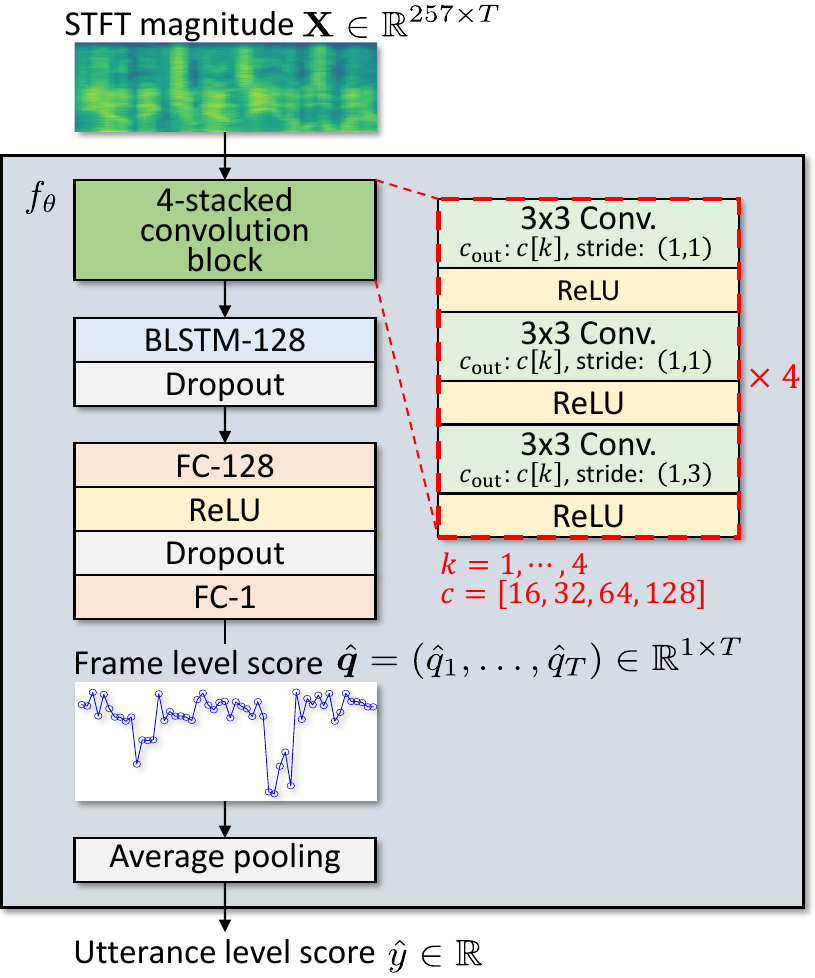} 
\vspace*{-0.45cm}
 \caption{Architecture of MOSNet. The symbol $c_{\rm out}$ corresponds to the number of output channels.}
 \label{fig:mosnet1}
 \vspace*{-0.7cm}
\end{figure}

\vspace*{-0.2cm}
MOSNet~\cite{lo2019mosnet} is the pioneering automatic SQA model for VC speech samples.
The architecture of MOSNet is shown in Fig.~\ref{fig:mosnet1}.
The input-output relationship of the model can be expressed as
$\hat{y}=f_{\theta}(\mathbf{X})$,
where $\theta$ is the set of all network parameters to be trained, and $\mathbf{X}$ is the short-time Fourier transform (STFT) magnitude spectrogram of a speech sample. First, $\mathbf{X}$ is fed into the 2D convolutional neural networks (CNNs) to obtain a intermediate representation that captures the local time-frequency structure. Next, a valuable feature is extracted by inputting the intermediate representation into a bidirectional long short-term memory (BLSTM). The frame-level score $\hat{\bm{q}}=(\hat{q}_{1},\ldots,\hat{q}_{T})$ are predicted by passing the valuable feature through two fully-connected (FC) layers, where $T$ denotes the number of frames. Finally, the utterance-level score $\hat{y}$ is obtained by applying the average pooling to $\hat{\bm{q}}$.

For training of $\theta$, the pair $(\mathbf{X}_{i}, y_{i})$ of the STFT magnitude spectrogram $\mathbf{X}_{i}$ and the actual MOS $y_{i}=1/N_{i}\sum_{n=1}^{N_{i}}s_{i,n}$ of speech sample $i$ is used, where $\bm{s}_{i}=(s_{i,1},\ldots,s_{i,N_{i}})$ are the set of all ACR-format ratings assigned to speech sample $i$, and  $N_{i}$ is the number of the ratings. The parameters $\theta$ are updated by backpropagation with the loss
\begin{align}
\label{eq:trainloss}
    \mathcal{L}=\left(\hat{y}_{i}-y_{i}\right)^2+\frac{\alpha}{T_i}\sum_{t}^{T_i}\left(\hat{q}_{i,t}-y_{i}\right)^2,
\end{align}
where $\hat{y}_{i}=f_{\theta}(\mathbf{X}_{i})$, $T_i$ is the number of frames of $\mathbf{X}_{i}$, $\hat{\bm{q}}_i=(\hat{q}_{i,1},\ldots,\hat{q}_{i,T_{i}})$ is the predicted frame-level MOS, and $\alpha$ is the weighting factor. The first term of $\mathcal{L}$ is the primary component to make the output of $f_{\theta}$ close to the actual MOS, while the second term serves as an auxiliary component to stabilize the predicted frame-level MOS $\hat{\bm{q}}_i$.
\vspace*{-0.2cm}
\subsection{Other neural-based SQA models}
\vspace*{-0.2cm}
To enhance the accuracy of MOS prediction models for VC speech samples, various approaches are explored.
SSL-MOS~\cite{cooper2022generalization} utilizes large pretrained audio models learned through SSL methods, such as wav2vec2.0~\cite{baevski2020wav2vec} and HuBERT~\cite{hsu2021hubert}.
NORESQA-MOS~\cite{manocha2022speech} employs the multi-task learning that encompasses the preference and relative rating tasks, both inspired by human’s ability to compare the quality of two speech signals.

Focusing on the variations in annotation patterns among listeners, several SQA models, such as MBNet~\cite{leng2021mbnet}, LDNet~\cite{huang2022ldnet}, and UTMOS~\cite{saeki2022utmos}, have been proposed to predict listener-dependent ratings instead of simply predicting MOS. The training of these models requires the listener's ID, which may not always be obtainable. Furthermore, in order to capture the rating tendencies of each listener, it is crucial for each listener to rate a reasonable number of speech samples. Our proposed method can enhance predictive performance without being bound by these constraints.
\vspace*{-0.2cm}
\subsection{Perceptual evaluation of TTS speech}
\vspace*{-0.2cm}
The perceptual evaluation of TTS speech samples by humans is considered to be influenced by several acoustic characteristics in a complex way, which has been analyzed from various perspectives.
In \cite{mayo2011listeners}, the study determines which acoustic characteristics of unit-selection TTS speech are most salient to listeners when evaluating the naturalness of such speech. The experimental result in \cite{clark2019evaluating} indicates that the MOS scores assigned to TTS speech paragraphs are lower than those assigned to TTS speech sentences, while no such difference is observed for clean speech samples. This result can be seen as consistent with our hypothesis.
\vspace*{-0.3cm}
\section{Subjective evaluation datasets}
In this section, we introduce the VCC2018~\cite{lorenzo_vcc2018} and BVCC~\cite{cooper2021voices} datasets. In these datasets, each speech sample was annotated by multiple listeners using the five point opinion scale from ``1: bad" to ``5: excellent."
\vspace*{-0.3cm}
\subsection{VCC2018 dataset~\cite{lorenzo_vcc2018}}
\vspace*{-0.2cm}
The VCC2018 dataset comprises 20,580 speech samples generated by VC methods submitted to VCC2018, all of which are spoken in US English.
Each speech sample was assessed by four listeners, except for 14 samples which received evaluations from three listeners, and one sample that was assessed by only two listeners. The number of listeners was 270.
\vspace*{-0.3cm}
\subsection{BVCC dataset~\cite{cooper2021voices}}
\vspace*{-0.2cm}
The BVCC dataset contains speech samples from the past Blizzard challenges, the voice conversion challenges, and TTS systems implemented in ESPNet\cite{hayashi2020espnet}. This dataset is divided into training and validation sets with sizes of 4,974 and 1,066 speech samples, respectively. This partition is chosen to hold out some unseen synthesis
systems, speakers, texts, and listeners in the validation set while matching the overall distributions of ratings as
closely as possible between the sets. Note that the validation set also serves the role of the test set. Each speech sample was rated by eight listeners. The number of listeners was 304.
\vspace*{-0.2cm}
\subsection{Analysis of datasets}
\vspace*{-0.2cm}
\begin{figure}[t]
 \vspace*{0cm}
 \centering
 \hspace*{-0cm}
\includegraphics[width=0.8\columnwidth]{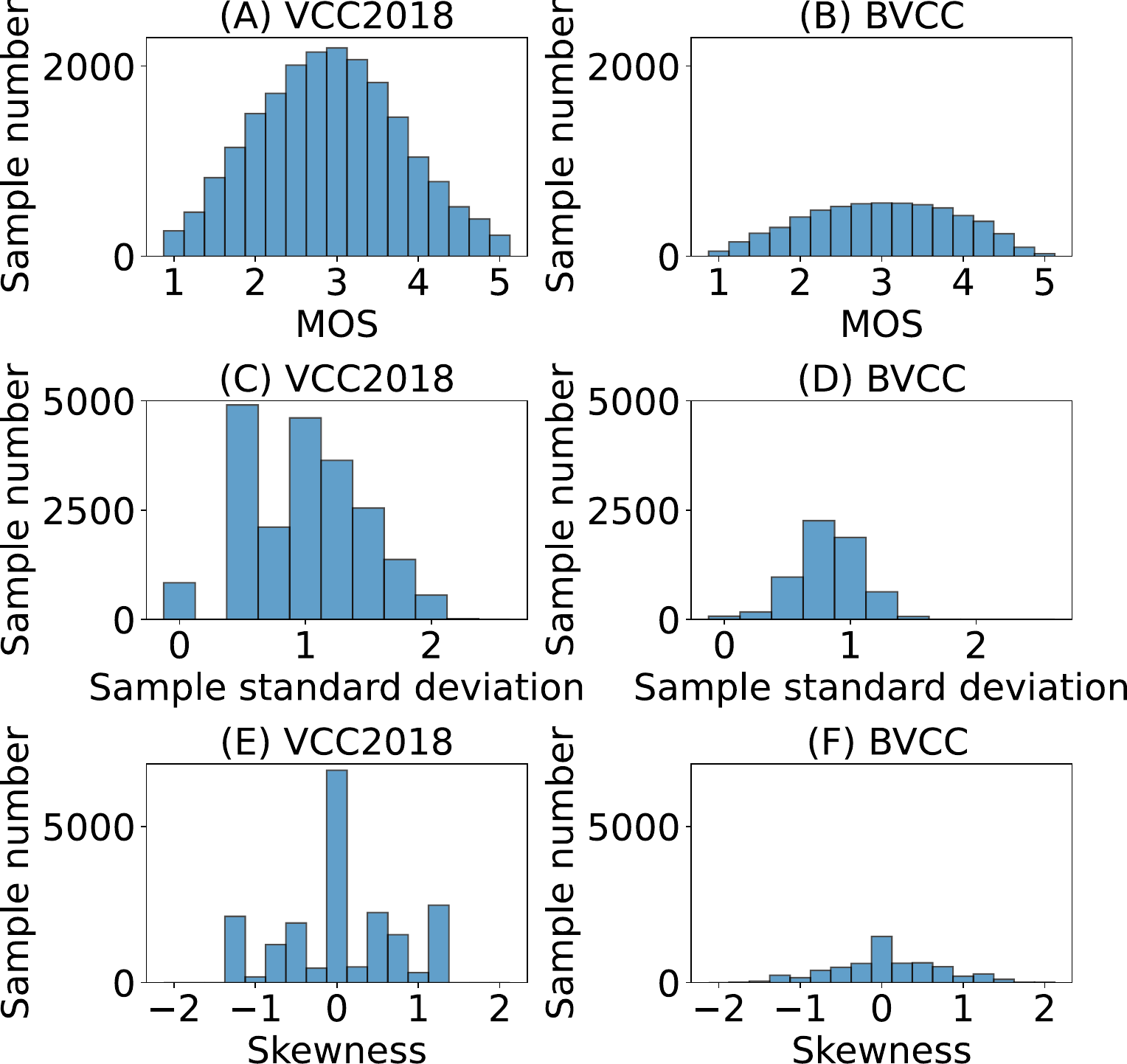}
\vspace*{-0.4cm}
\caption{Histograms of (A), (B) MOS, (C), (D) sample standard deviation, and  (E), (F) skewness of all ratings for each speech sample. (A), (C), and (E) are about the VCC2018 dataset, and (B), (D), and (F) are about the BVCC dataset.}
 \label{fig:hist_all}
 \vspace*{-0.7cm}
\end{figure}
\begin{table}[t]
\caption{The count of speech samples with skewness for each sign.}
\label{tab:num_skew}
\tabcolsep = 3pt
\centering
{
 \begin{tabular}{c||c|c|c|c} \bhline{1.5pt}
 Dataset&\begin{tabular}{c}
$+$
\end{tabular}&\begin{tabular}{c}
$-$
\end{tabular} &\begin{tabular}{c}
0
\end{tabular}&undefined \\ \bhline{1.5pt}
\begin{tabular}{c}VCC2018\end{tabular} & 7,128& 5,940 &6,675 & 837\\ \hline
\begin{tabular}{c}BVCC\end{tabular} & 2,767 & 2,312 &888 &73
\\\bhline{1.5pt}
 \end{tabular}
}
\vspace*{-0.6cm}
\end{table}
The histograms of the MOS (mean), sample standard deviation, and skewness of the distribution consisting of all ratings annotated to each speech sample are shown in Fig.~\ref{fig:hist_all}. Note that the speech samples for which skewness is undefined, which means the variance equals to 0, are not included in the histograms (E) and (F). The histograms indicate that the rating distributions for each speech sample are diverse in terms of MOS, variance, and skewness regarding the VCC2018 and BVCC datasets.

We display the count of speech samples with skewness for each sign in Table~\ref{tab:num_skew}. From this table, it can be observed that the number of samples with positive skewness is greater than that with negative skewness in each dataset. This indicates that the distribution of ratings annotated to each speech sample tends to be right-skewed.
This result may imply the validity of the hypothesis that the variance in ratings for each speech sample is primarily attributed to accidental inclusion of higher scores due to overlooking or neglecting poor quality speech segments, since such accidental higher scores skew the rating distribution to the right.
\vspace*{-0.3cm}
\section{Proposed representative value of subjective speech quality}
\vspace*{-0.2cm}
\subsection{$N_{\rm low}$-MOS}
\vspace*{-0.2cm}
From our hypothesis that the variance in ratings for each speech sample is primarily attributed to accidental inclusion of higher scores, lower ratings are considered to be more reliable than higher ones. Focusing on this, we propose the more intrinsic representative value of subjective speech quality $N$-lowest MOS (or $N_{\rm low}$-MOS for short), the mean of the $N$-lowest opinion scores.

For speech sample $i$, let $s_{i, 1}, s_{i,2}, ..., s_{i,N_{i}^{\rm (all)}}$ be the sequence of $N_{i}^{\rm (all)}$ ACR-format ratings arranged in ascending order. The $N$-lowest MOS $y_{i}^{\rm (low)}$ for the speech sample $i$ is defined as 
\vspace*{-0.2cm}
\begin{align}
    y_{i}^{\rm (low)}=\frac{1}{N}\sum_{n=1}^{N}s_{i,n}.
\end{align}
$1$-lowest MOS and $N_{i}^{\rm (all)}$-lowest MOS correspond to the minimum value and the regular MOS, respectively.
$N_{\rm low}$-MOS does not represent the MOS score itself, but $N_{\rm low}$-MOS also serves as a usable representative value of subjective speech quality when comparing subjective speech quality of processed speech samples. The comparison is fair among the audio samples in cases where the numbers of ratings annotated to each speech sample are all the same.

For the experiment section, we define $N$-highest MOS (or $N_{\rm high}$-MOS for short). $N$-highest MOS $y_{i}^{\rm (high)}$ is defined as the average of the $N$ highest ratings:
\vspace*{-0.2cm}
\begin{align}
    y_{i}^{\rm (high)}=\frac{1}{N}\sum_{n=1}^{N}s_{i,N_{i}^{\rm (all)}+1-n}.
\end{align}
We also define $(N^{\rm (low)},N^{\rm (high)})$-central MOS. $(N^{\rm (low)},N^{\rm (high)})$-central MOS $y_{i}^{\rm (central)}$ for the speech sample $i$ is defined as the average of all the ratings except $N^{\rm (low)}$ lowest and $N^{\rm (high)}$ highest ratings:
\vspace*{-0.2cm}
\begin{align}
    y_{i}^{\rm (central)}=\frac{1}{N_{i}^{\rm (all)}-N^{\rm (low)}-N^{\rm (high)}}\sum_{n=N^{\rm (low)}+1}^{N_{i}^{\rm (all)}-N^{\rm (high)}}s_{i,n}.
\end{align}
$(N^{\rm (low)},N^{\rm (high)})$-central MOS can correspond to the median by appropriately choosing $N^{\rm (low)}$ and $N^{\rm (high)}$.

\begin{figure}[t]
 \vspace*{0cm}
 \centering
 \vspace*{-0cm}
\includegraphics[width=0.6\columnwidth]{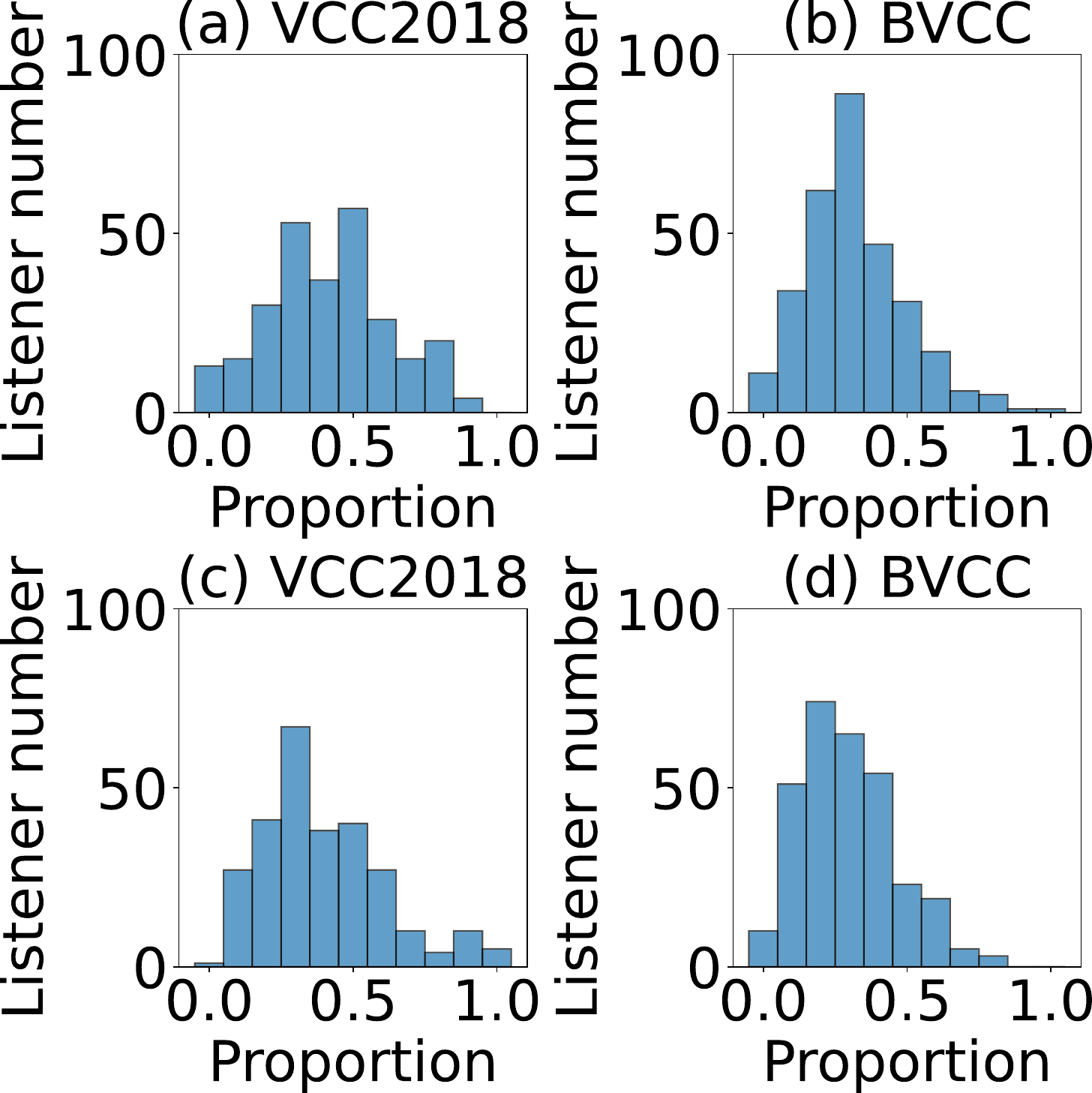}
\vspace*{-0.45cm}
\caption{Histograms of (a) and (b) the proportion of ratings used as $1$-lowest MOS for each listener, (c) and (d) the proportion of ratings used as $1$-highest MOS for each listener. (a) and (c) are about the VCC2018 dataset, and (b) and (d) are about the BVCC dataset.}
 \label{fig:hist_ratio}
 \vspace*{-0.5cm}
\end{figure}

For small values of $N$, only a very limited subset of ratings may be used in calculating $N_{\rm low}$-MOS or $N_{\rm high}$-MOS, potentially reflecting the opinions of only a small number of listeners. We show the histograms of the proportion of ratings used as $1$-lowest MOS or $1$-highest MOS for each listener in Fig.~\ref{fig:hist_ratio}. It can be confirmed that $1$-lowest MOS and $1$-highest MOS are not dependent on a small number of listeners, who may be uncooperative.
\vspace*{-0.2cm}
\subsection{Applying to prediction models}
\vspace*{-0.2cm}
From the above hypothesis, the application of MOS to subjective speech quality predictor models is consider to set the upper limit on the prediction performance. To relax the limit, we propose applying $N_{\rm low}$-MOS to training of subjective speech quality prediction models.
This idea is applicable to any model to be trained using MOS, including MOSNet, SSL-MOS, and NORESQA-MOS. If the use $N_{\rm low}$-MOS instead of MOS improves LCC and SRCC, it can be said that the prediction model becomes more useful as a comparator of speech samples.
\vspace*{-0.4cm}
\section{Experiment}
\vspace*{-0.3cm}
\subsection{Experimentally condition}
\vspace*{-0.2cm}
We compared the prediction performance when applying each representative value of subjective speech quality to MOSNet~\cite{lo2019mosnet}.
We used the VCC2018~\cite{lorenzo_vcc2018}  and BVCC~\cite{cooper2021voices} datasets.
In the VCC2018 dataset experiment,
we compared MOS, $N$-lowest MOS, and $N$-highest MOS $(N=1, 2, 3)$. In the VCC2018 dataset experiment, we divided the speech samples into training, validation, and test sets with sizes of 13,580, 3,000, and 4,000, respectively. Note that all listeners for validation and test data are most likely seen listeners.
In the BVCC dataset experiment,
we compared MOS, $N$-lowest MOS, $N$-highest MOS $(N=1, 2, \ldots, 7)$ and $(N^{\rm (low)},N^{\rm (high)})$-central MOS $\big((N^{\rm (low)},N^{\rm (high)})=(2,1), (1,2), (3,3)\big)$, where $(3,3)$-central MOS corresponds to the median.

The sampling frequency was 16~kHz. Hamming window was used as a window function for STFT. The window length and shift length were 32~ms and 16~ms, respectively. For MOSNet, $\alpha$ was set to $1$. The model was trained using the Adam optimizer~\cite{kingma2014adam} with a learning rate of 0.0001. The batch size was 32. We applied early stopping
based on the validation loss with five epochs patience. The dropout rate~\cite{srivastava2014dropout} was set to 0.3. We performed eight trials by randomizing the initial values of network parameters $\theta$ using a standard normal distribution. In the VCC2018 dataset experiment, we also randomized the division of the dataset into training, validation, and test sets. 
\vspace*{-0.4cm}
\subsection{Results}

\begin{table}[t]
\caption{The mean and sample standard deviation of utterance-level MSE, LCC and SRCC results of the experiment on the VCC2018 dataset over eight trials. Bold indicates superiority over regular MOS. Note that the MSE is included for reference purposes only and is not suitable for the comparison between MOS, $N_{\rm low}$-MOS, and $N_{\rm high}$-MOS, as the distribution of representative values in the test sets differs for each representative value.}
\label{tab:result_VCC2018}
\tabcolsep = 3pt
\centering
{
\footnotesize{
 \begin{tabular}{c|c||c|c|c} \bhline{1.5pt}
 \begin{tabular}{c}
Representative\\value
\end{tabular}&\begin{tabular}{c}
$N$
\end{tabular} &\begin{tabular}{c}
MSE 
\end{tabular} &\begin{tabular}{c}
LCC $\uparrow$
\end{tabular} &\begin{tabular}{c}
SRCC $\uparrow$
\end{tabular} \\ \bhline{1.5pt}
Regular MOS &N/A& 0.500$\pm$0.028 & 0.614$\pm$0.019 & 0.576$\pm$0.018 \\ \hline
\multirow{3}{*}{$N_{\rm low}$-MOS}& 1 & 0.677$\pm$0.026 & 0.515$\pm$0.016 & 0.468$\pm$0.016 \\ \cline{2-5}
 & 2& 0.596$\pm$0.029 & 0.594$\pm$0.013 & 0.558$\pm$0.010 \\ \cline{2-5}
 & 3& 0.549$\pm$0.014 & \textbf{0.622$\pm$0.014} & \textbf{0.582$\pm$0.016} \\ \hline
 \multirow{3}{*}{$N_{\rm high}$-MOS}&1& 0.854$\pm$0.071 & 0.307$\pm$0.184 & 0.284$\pm$0.173 \\ \cline{2-5}
 &2& 0.668$\pm$0.018 & 0.525$\pm$0.019 & 0.503$\pm$0.018 \\ \cline{2-5}
&3& 0.598$\pm$0.030 & 0.572$\pm$0.023 & 0.545$\pm$0.020  
\\\bhline{1.5pt}
 \end{tabular}
}}
\vspace*{-0.7cm}
\end{table}

\begin{table}[t]
\caption{The mean and sample standard deviation of utterance-level MSE, LCC and SRCC results of the experiment on the BVCC dataset over eight trials. Bold indicates superiority over MOS.}
\label{tab:result_BVCC}
\tabcolsep = 3pt
\centering
{
\footnotesize{
 \begin{tabular}{c|c||c|c|c} \bhline{1.5pt}
 \begin{tabular}{c}
Representative\\value
\end{tabular}&\begin{tabular}{c}

$N$
\end{tabular} &\begin{tabular}{c}
MSE 
\end{tabular} &\begin{tabular}{c}
LCC $\uparrow$
\end{tabular} &\begin{tabular}{c}
SRCC $\uparrow$
\end{tabular} \\ \bhline{1.5pt}
Regular MOS &N/A& 0.608$\pm$0.067 & 0.540$\pm$0.081 & 0.541$\pm$0.070 \\ \hline
\multirow{7}{*}{$N_{\rm low}$-MOS}& 1 & 0.587$\pm$0.030 & 0.505$\pm$0.040 & 0.494$\pm$0.049 \\ \cline{2-5}
 & 2& 0.566$\pm$0.071 & 0.527$\pm$0.089 & 0.522$\pm$0.091 \\ \cline{2-5}
 & 3& 0.529$\pm$0.063 &\textbf{ 0.583$\pm$0.064} & \textbf{0.579$\pm$0.065} \\ \cline{2-5}
 & 4& 0.525$\pm$0.049 & \textbf{0.607$\pm$0.054} & \textbf{0.596$\pm$0.061} \\ \cline{2-5}
 & 5& 0.587$\pm$0.098 & \textbf{0.571$\pm$0.094}
 & \textbf{0.568$\pm$0.083} \\ \cline{2-5}
 & 6& 0.553$\pm$0.075 & \textbf{0.612$\pm$0.080} & \textbf{0.606$\pm$0.075} \\ \cline{2-5}
 & 7&  0.585$\pm$0.136 & \textbf{0.581$\pm$0.127} & \textbf{0.582$\pm$0.126} \\ \hline
 \multirow{7}{*}{$N_{\rm high}$-MOS}&1& 0.844$\pm$0.085  & 0.375$\pm$0.093 & 0.373$\pm$0.089 \\ \cline{2-5}
 &2& 0.772$\pm$0.090  &  0.446$\pm$0.085 & 0.452$\pm$0.078 \\ \cline{2-5}
&3& 0.720$\pm$0.120  &  0.522$\pm$0.108 & 0.519$\pm$0.100 \\ \cline{2-5}
&4& 0.744$\pm$0.098  & 0.496$\pm$0.089 & 0.498$\pm$0.080 \\ \cline{2-5}
 &5& 0.699$\pm$0.094  & 0.530$\pm$0.092 & 0.527$\pm$0.077 \\ \cline{2-5}
 &6& 0.632$\pm$0.122  & \textbf{0.576$\pm$0.115} & \textbf{0.570$\pm$0.109} \\ \cline{2-5}
&  7& 0.644$\pm$0.121 & 0.538$\pm$0.113 & \textbf{0.543$\pm$0.104} \\ \hline
 \multirow{3}{*}{\begin{tabular}{c}$(N^{\rm (low)},N^{\rm (high)})$\\-central MOS\end{tabular}}&(2,1)& 0.701$\pm$0.115 & 0.539$\pm$0.104 & 0.541$\pm$0.090 \\ \cline{2-5}        
 &(1,2)& 0.641$\pm$0.099 & \textbf{0.575$\pm$0.096} & \textbf{0.573$\pm$0.082} \\ \cline{2-5}        
& (3,3)& 0.770$\pm$0.115 & 0.539$\pm$0.099 & \textbf{0.544$\pm$0.084}      
\\\bhline{1.5pt}
 \end{tabular}
}}
\vspace*{-0.75cm}
\end{table}

\vspace*{-0.2cm}
We show the averaged MSE, LCC and SRCC results of the experiments on the VCC2018 and BVCC datasets in Tables \ref{tab:result_VCC2018} and \ref{tab:result_BVCC}, respectively. In both tables, it can be observed that the $N$-lowest MOS achieves higher LCC and SRCC than the $N$-highest MOS for all values of $N$. This result suggests that lower ratings are more reliable than higher ones, and $N_{\rm low}$-MOS represents subjective speech quality better than $N_{\rm high}$-MOS. In Table \ref{tab:result_VCC2018}, $3$-lowest MOS achieves higher LCC and SRCC performance than MOS, while $N_{\rm high}$-MOS does not improve the performance for all degrees $N$. In Table \ref{tab:result_BVCC}, $N$-lowest MOS $(N=3,4,5,6,7)$ achieve higher LCC and SRCC performance than regular MOS, while $N_{\rm high}$-MOS does not improve the performance for almost all degrees $N$. These results indicate that, though $N_{\rm high}$-MOS is not so attractive, $N_{\rm low}$-MOS tends to represent subjective speech quality better than regular MOS.

From Tables \ref{tab:result_VCC2018} and \ref{tab:result_BVCC}, it can be observed that, the averaged LCC and SRCC obtained when using $N$-lowest MOS $(N=1,2)$ are lower compared to when using MOS. This result is considered to be influenced by some degree of human error and individual preferences in the $2$-lowest ratings, which is corresponding to the better prediction performance of $6$-highest MOS than regular MOS. This result emphasizes the effectiveness of $N_{\rm low}$-MOS for an adequate value of $N$, since it has properties intermediate between $1$-lowest MOS (i.e., the minimum value) and regular MOS.

 In Table~\ref{tab:result_BVCC}, all types of $(N^{\rm (low)},N^{\rm (high)})$-central MOS give lower LCC and SRCC compared to when using $6$-lowest MOS. It reveals that the strategy of blindly excluding outliers from both ends does not necessarily yield the best result.
 \vspace*{-0.45cm}
\section{Conclusion}
\vspace*{-0.35cm}
We analyzed the VCC2018 and BVCC datasets, and for the first time discovered that the distribution of all ratings for each speech sample tends to skew to the right. We believe that our findings suggest the validity of our hypothesis that, when humans subjectively rate speech, they tend to assign more weight to low-quality speech segments and the variance in ratings for each speech sample is mainly due to accidental assignment of higher scores when overlooking the poor quality segments. We proposed applying $N_{\rm low}$-MOS to training of subjective speech quality prediction models, which relaxes the prediction performance limit due to the use of regular MOS. We experimentally confirmed the superiority of $N_{\rm low}$-MOS over regular MOS as the training data of the MOSNet architecture, which suggests that $N_{\rm low}$-MOS is a more intrinsic representative value of subjective speech quality and more useful when comparing speech quality.
This work can motivate researchers interested in SQA to conduct a more in-depth analysis of subjective evaluation datasets and further explore intrinsic representative values for subjective speech quality.

\bibliographystyle{IEEEbib}
\bibliography{refs}

\end{document}